\documentclass[onecolumn,12pt]{revtex4-2}\pdfoutput=1

\usepackage{color}
\usepackage{mathtools}

\usepackage{hyperref}
\hypersetup{ colorlinks,
linkcolor=blue,
filecolor=blue,
urlcolor=blue,
citecolor=blue}

\usepackage{amsmath}
\usepackage{amssymb}
\usepackage{bbding}
\usepackage{wasysym}
\usepackage{mathtools}
\usepackage{xcolor}
\usepackage{geometry}
\begin{document}

\title{Topology-induced modifications in the critical behavior of the Yaldram–Khan catalytic reaction model}
\author{Paulo F. Gomes\textsuperscript{1,2}, Henrique A. Fernandes\textsuperscript{1}, Roberto da Silva\textsuperscript{3}}
\affiliation{1 - Grupo de Redes Complexas Aplicadas de Jataí, Universidade Federal de Jata{\'i}, BR 364, km 192, 3800 - CEP 75801-615, Jata{\'i}, Goi{\'a}s, Brazil\\
2 - Faculdade de Ciências e Tecnologia, Universidade Federal de Goi{\'a}s, Estrada Municipal, Bairro Fazenda Santo Antônio, CEP 74971-451, Aparecida de Goiânia, Goi{\'a}s - Brasil\\
3 - Instituto de F{\'i}sica, Universidade Federal do Rio Grande do Sul, Av. Bento Gon{\c{c}}alves, 9500 - CEP 91501-970, Porto Alegre, Rio Grande do Sul, Brazil}

\begin{abstract}
In this work, we investigated how the use of complex networks as catalytic surfaces can affect the phase diagram of the Yaldram-Khan model, as well as how the order of the phase transitions present in the seminal work behaves when the randomness is added to the model. The study was conducted by taking into consideration two well-known random networks, the Erdős-Rényi network (ERN), with its long-range randomness, and the random geometric graph (RGG), with its spatially constrained randomness. We perform extensive steady-state Monte Carlo simulations for $r_{\text{NO}}=1$, the NO dissociation rate, and show the behavior of the reactive window as function of the average degree of the networks. Our results also show that, different from the ERN, which preserves the nature of the phase transitions of the original model for all considered average degrees, the RGG seems to have two second-order phase transitions for small values of average degree.

\end{abstract}

\maketitle

\section{Introduction}

Theoretical approaches to reactions on catalytic surfaces — accounting for processes such as adsorption, desorption, diffusion, and interactions between adsorbed species -- play a crucial role in understanding and interpreting experimental results in surface science and heterogeneous catalysis~\cite{norskov2008density, hammer1995theoretical, niemantsverdriet2007spectroscopy}. Among the various theoretical frameworks, statistical mechanics -- particularly its branch dedicated to nonequilibrium systems described by stochastic rules or transition probabilities, rather than by a defined Hamiltonian~\cite{Hinrichsen2000, Henkel} — offers powerful tools for modeling such phenomena. Its success is evidenced by the wide range of different applications, including epidemic-like models~\cite{TomeJPA, SilvaJSTAT}, voter models~\cite{SilvaVoter}, and other related approaches, all of which provide valuable insights into the critical behavior of these systems.

The application of statistical mechanics to model catalytic surface reactions was pioneered by Ziff, Gulari, and Barshad (ZGB)~\cite{ziff1986}, who introduced a model that captures the formation of carbon dioxide ($\text{CO}_2$) as a result of the reaction between carbon monoxide (CO) molecules and oxygen (O) atoms adsorbed on a catalytic surface. Today, the well-known ZGB model stands as one of the most significant models in the field of catalytic surface reactions. Its importance stems from its simplicity and its remarkable ability to exhibit both continuous (second-order) and discontinuous (first-order) phase transitions, making it a compelling framework for exploring the dynamics and critical phenomena associated with surface reactions.

After its advent, steady-state and time-dependent Monte Carlo (MC) simulations have yielded several interesting results. For instance, the introduction of desorption, diffusion, and other processes has led to new phase transitions not present in the original model (see, for example, Refs. \cite{marro1999, tome1993, RdasilvadifusionZGB2018, Fernandes2016primeiroZGB, Fernandesdesopsion, Fernandes2025}). Another interesting model, devised by K. Yaldram and M. A. Khan, takes into consideration the reactions presented in the ZGB model and, in addition, incorporates additional processes to build a more complex model, today named the YK model \cite{Yaldram1991}. They considered the adsorption of nitric oxide (NO) molecules on the catalytic surface, resulting in the desorption of nitrogen ($\text{N}_{2}$) molecules and the formation of a stable surface of oxygen atoms. Subsequently, these atoms are consumed by carbon monoxide (CO) molecules, leading to their oxidation to carbon dioxide ($\text{CO}_{2}$) molecules, which also leave the surface. The overall simplified surface reaction can be expressed as
\begin{equation}
\text{NO}+\text{CO}{\longrightarrow} \text{N}_{2} + \text{CO}_{2}
\label{Eq:General_Equation}
\end{equation}

Yaldram and Khan \cite{Yaldram1991} were the first to propose a Monte Carlo simulation model for the NO-CO catalytic reaction, decomposing the Eq. (\ref{Eq:General_Equation}) into seven distinct steps. Their results demonstrated that there is no reactive state for regular square lattices. However, in a hexagonal lattice, the steady-state CO concentration exhibits two key features: a critical value (associated with a continuous phase transition) and a higher value corresponding to a first-order (discontinuous) phase transition.

The model has been studied by several authors, with different approaches and methods \cite{Khan2002, Loscar2003, Avalos2006}. In Ref. \cite{Huiyun1996}, a cellular automaton approach was considered along with mean-field analysis to show that, under certain circumstances, there exists an active phase for regular square lattices. The pair approximation was derived in Ref. \cite{Dickman1999} and show that a steady reactive window is found for both triangular and square lattices. The diffusion of particles on the catalytic surface was addressed in Refs. \cite{Khan1994, Aida1999, Luque2004} and impurities in Refs. \cite{Lorenz2002, Ahmad2007}. The desorption of some particles is considered in Refs. \cite{Meng1994, Diaz2018, Hernandez2022} in order to analyze the effect of temperature on the system. In some of those works, the inclusion of impurities, diffusion, or desorption has proven to be sufficient to allow a reactive steady-state on regular square lattices. However, to the best of our knowledge, the study of the behavior of the phase transitions when the catalytic reactions occur on random networks has not yet been carried out, as has been done for the ZGB model \cite{Vilela2020}.

In order to address this question, we decided to study the YK model on two different random networks: the Erdös-Rényi network (ERN) and the random geometric graph (RGG) whereas in realistic scenarios, surfaces are influenced by additional stochastic effects, which exhibit structural randomness. In this study, we examine how this randomness alters the phase transitions of the model and quantitatively assess their dependence on the average connectivity of the networks under investigation.

The paper is structured as follows: In Section \ref{sec_model}, we outline the YK model and the properties of the graphs used in our study. Section \ref{sec_results} begins with a pedagogical revisit of the hexagonal lattice case before exploring complex networks. Next, we analyse the phase transitions as function of the average degree and give some clues about the order of these transitions for both networks. In that section, we also show how the reactive window, defined as the difference between the two phase-transition points separating the absorbing state from the active phase, varies with the network parameters. Finally, our conclusions are summarized in Section \ref{sec_conclusions}.


\section{The model and the catalytic surfaces} \label{sec_model}


\subsection{YK model} \label{subsec_ykmodel}

The model proposed by Yaldram and Khan has attracted a lot of attention since their seminal work due to its simplicity, rich phase diagram presenting both continuous and discontinuous phase transitions, and prominent possibility of technological applications when considering the study of ways to overcome the problems created by polluting gases resulting from the consumption of fossil fuels. As named after its proposition, the YK model can be seen as processes involving both adsorption, desorption, and dissociation of molecules on/from/over a catalytic surface. The reactions can be summarized in the following equations:

\begin{equation}
\text{CO}(g)+V \overset{y}{\longrightarrow} \text{CO}(a), \label{eq:co_ad}
\end{equation}
\begin{equation}
\text{NO}(g)+V \overset{1-y}{\longrightarrow} \text{NO}(a), \label{eq:no_ad}
\end{equation}
\begin{equation}
\text{NO}(g)+2V \overset{r_{\text{NO}}}{\longrightarrow} \text{N}(a) + \text{O}(a), \label{eq:n_o_ad}
\end{equation}
\begin{equation}
\text{NO}(a)+\text{N}(a) \overset{1}{\longrightarrow} \text{N}_2(g)+\text{O}(a)+V, \label{eq:n2_g_o_ad}
\end{equation}
\begin{equation}
\text{N}(a)+\text{N}(a) \overset{1}{\longrightarrow} \text{N}_2(g)+2V, \label{eq:n2_g}
\end{equation}
\begin{equation}
\text{CO}(a)+\text{O}(a)\overset{1}{\longrightarrow} \text{CO}_{2}(g) + 2V, \label{eq:co2_des}
\end{equation}
where N and O are nitrogen and oxygen atoms, CO and $\text{CO}_2$ are respectively, carbon monoxide and carbon dioxide molecules, and the vacant sites on the surface are represented by the letter $V$. The model possesses two parameters: the adsorption rate of CO molecules, $y$, and the dissociation rate of NO molecules, $r_{NO}$. Equation (\ref{eq:co_ad}) represents the process in which, with a rate $y$, a CO molecule in the gas $(g)$ phase is chosen to impinge the surface, being absorbed $(a)$ if it hit a vacant site $V$. On the other hand, as shown in Eq. (\ref{eq:no_ad}), with a rate $1-y$ a NO molecule in the gas phase is chosen to collide with the surface. In that case, there are two possibilities depending on the dissociation rate, $r_{\text{NO}}$, the probability the NO molecule has to dissociate into N and O atoms. As shown in Eq. (\ref{eq:n_o_ad}), with that rate, the dissociation occurs and the atoms are adsorbed on the surface whenever two neighboring sites, chosen at random, are vacant. Otherwise, the NO molecule does not dissociate and it is adsorbed on the surface if the chosen site is vacant (see Eq. (\ref{eq:no_ad})). During any adsorption process, if either chosen site is occupied by an atom/molecule, the trial ends and the CO or NO molecule returns to the gas phase. On the other hand, whenever an adsorption process is successful, the neighborhood of the newly adsorbed atom/molecule is randomly checked. When an NO molecule is adsorbed and there is, at least, one neighboring N atom, a $\text{N}_2$ molecule is formed and desorbs from the surface, leaving behind one O atom at the site where was the NO molecule and a vacant site. As can be seen, this reaction is represented by Eq. (\ref{eq:n2_g_o_ad}) and, whenever it occurs, the N atom left on the surface can also react with another neighboring N atom, as shown in Eq. (\ref{eq:n2_g}), producing a $\text{N}_2$ molecule which desorbs from the surface leaving two empty sites. Finally, when a CO molecule is adsorbed on the surface, the neighboring sites are randomly checked for O atoms. As shown in Eq. (\ref{eq:co2_des}), if an O atom is found, a $\text{CO}_2$ molecule is immediately formed and desorbed from the surface, leaving two vacant sites on it. 

After the system reaches the steady state, the densities of the adsorbed species, as well as the density of vacant sites, are computed as follows:
\begin{equation}
\rho _{\xi }=\frac{N_{_{\xi }}}{N},
    \label{eq:densities}
\end{equation}
where $N$ is the total number of sites on the catalytic surface, and $N_{\xi}$ is the number of entities corresponding to species $\xi=V$, O, CO, $\text{N}_2$, and $\text{CO}_2$. $N_\text{O}$ and $N_{CO}$ represent the number of adsorbed atoms/molecules on the surface, and $\text{N}_2$ and $\text{CO}_2$ represent the number of molecules that exist only in the gas phase and are produced only in the steady active phase.

\begin{figure} [!htbp] 
\begin{center}
\includegraphics[width = 6.0 in]{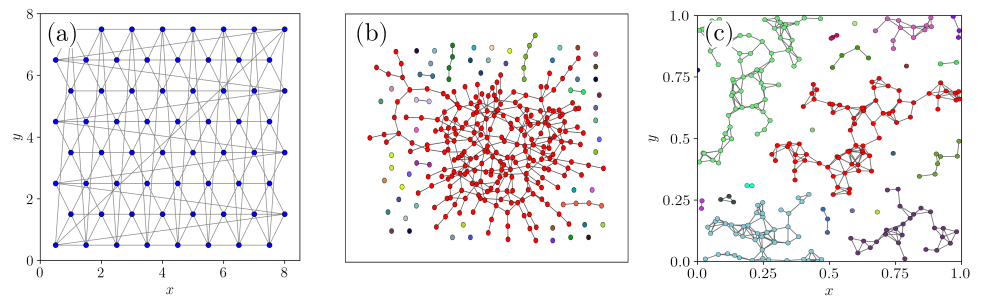}
\caption{Graphical representation of networks. (a) Regular hexagonal lattice with $N=64$ nodes and periodic boundary conditions. (b) Erdös-Rényi Network (ERN) with $N=300$ nodes and average degree $\mu = 2.0$. (b) RGG with $N=256$ nodes, radius $r = 0.07$, average degree $K \simeq 3.73$, and square size $L=1.0$. Each color represents a distinct component. The largest component is depicted in red.}
\label{fig_YK_s56g6}
\end{center}
\end{figure}

As designed by Yaldram and Khan \cite{Yaldram1991}, the model does not present a steady reactive state for regular square lattices. However, they found phase diagrams with continuous and discontinuous phase transitions separating absorbing phases from a steady active one, with production of $\text{CO}_2$ and $\text{N}_2$ molecules, for regular hexagonal lattices (see Fig. \ref{fig_YK_s56g6}(a)). 

As stated in the previous section, most previous works have focused solely on regular lattices, such as square and hexagonal structures, to model the catalytic surface. However, real surfaces can be more complex, and random effects play an important role considering the possibility of future realizations of experiments beyond of computer ones here performed. In the following, we present the random surfaces considered in this work, where the catalytic reactions occur.

\subsection{Random networks}

The present study aims to analyze the influence of random networks and its tunable topological properties on the phase diagram of the YK model. A network or a graph comprises a collection of vertices interconnected by a set of edges \cite{Newman2010,BarabasiREVMOD}. Each edge or connection links a pair of vertices, making them neighbors, with the number of vertices $N$ defining the network size and the set of edges determining its connectivity. In addition, when studying random networks, some parameters are commonly used, such as the degree $k_i$ which is defined as the number of neighbors of a given vertex $i$ and the average degree of the network, $m$, which is given by 
\begin{equation}
    m = \frac{1}{N}\sum_i^N k_i. \label{avergadlkj}
\end{equation}
In this work, terms commonly used in network science, such as vertices/nodes, refer, respectively, to sites on the catalytic surface and atoms or molecules either adsorbed on it or present in the gas phase. An ``empty or vacant site'' signifies the absence of an adsorbed particle on that site.

We took into consideration two well-known random networks, the ERN and the RGG, which are widely known and used in various areas of knowledge \cite{Gilbert1961,Seshadhri2012,Moore2015,Saeedian2019, Newman2002,Reia2019a,Reia2020a,Gomes2022,Franco2024}. The ERN \cite{Solomonoff1951, Erdos1959, Barabasi2016} is a random graph composed of $N$ vertices, with each pair connected independently with probability $\beta$, as illustrated in Fig. \ref{fig_YK_s56g6}(b). The control parameter of the network is the average degree, defined by construction as 
\begin{equation}
\mu = \beta (N - 1). \label{muasfunctionofbeta}
\end{equation}
If we generate many random samples of the ERN, the resulting $K$ values will, by construction, follow a binomial distribution. In the limit of $\beta \ll 1$ and $N \gg 1$, this distribution approaches a Poisson one, and more generally, for sufficiently large $N$, it can be approximated by a Gaussian distribution centered at $\mu$. Thus, the control parameter for the ERN is typically taken to be $\mu$ rather than the connection probability $\beta$.

On the other hand, the RGG \cite{Dall2002} is constructed by placing $N$ nodes at random in a square box of size $L$ in the $xy$ cartesian plane, as shown in Fig. \ref{fig_YK_s56g6}(c). The criterion for defining connectivity is the following: two sites are connected, that is, they are neighbors, if and only if the euclidean distance between them is less than a radius $r$, the control parameter of the network. We maintain the superficial density of sites $z = N/L^2$ constant in order to eliminate its effect when the size $N$ of the network is adjusted. So, we choose $L=\sqrt{N/z}$. Since the effective area of one site is $1/z$, we adopt the linear size $1/\sqrt{z}$ as the spatial scale to measure all distances in our study \cite{Gomes2019}. In this way, the network properties do not depend on the linear size of the square box and are uniquely defined by the radius $r$ measured in units of $1/\sqrt{z}$. As any constant value for $z$ works, we set $z=1.0$ so that the linear size of the square box becomes $\sqrt{N}$, and the radius $r$ is a dimensionless number. Therefore, the calculated average degree $K$ (see Eq. \ref{avergadlkj}) of the network varies from 0, for $r=0$, to $N-1$, when $r \geq L \sqrt{2} /2$ (half the diagonal of the square).

\section{Results \label{sec_results}}

In this section, we present our main results for the YK model simulated on the two complex networks here proposed. The estimates have been obtained by means of steady-state MC simulations when the dissociation rate, $r_{\text{NO}}$, is equal to one, i.e., every NO molecule in the gas phase which impinges the surface dissociates into a N and O atoms before being adsorbed on it (see Eq. (\ref{eq:n_o_ad})). The computational details are provided in Appendix A of this manuscript.  

However, before looking into the influence of the ERN and RGG on phase diagram of the model, we carried out simulations on the hexagonal network (HN) in order to define some important parameters to the MC method, such as the lattice size, $N$ (the number of sites/nodes), $\tau$, which is the number of MC steps discarded at the beginning of the simulations to ensure that the system reached the steady-state regime before computing the average densities $\rho_{\xi}$ (see Eq. (\ref{eq:densities})), and the number $S$ of MC steps considered to obtain these mean values. 

\subsection{Calibration}

\begin{figure} [!htbp]
\begin{center}
\includegraphics[width = 3.2 in]{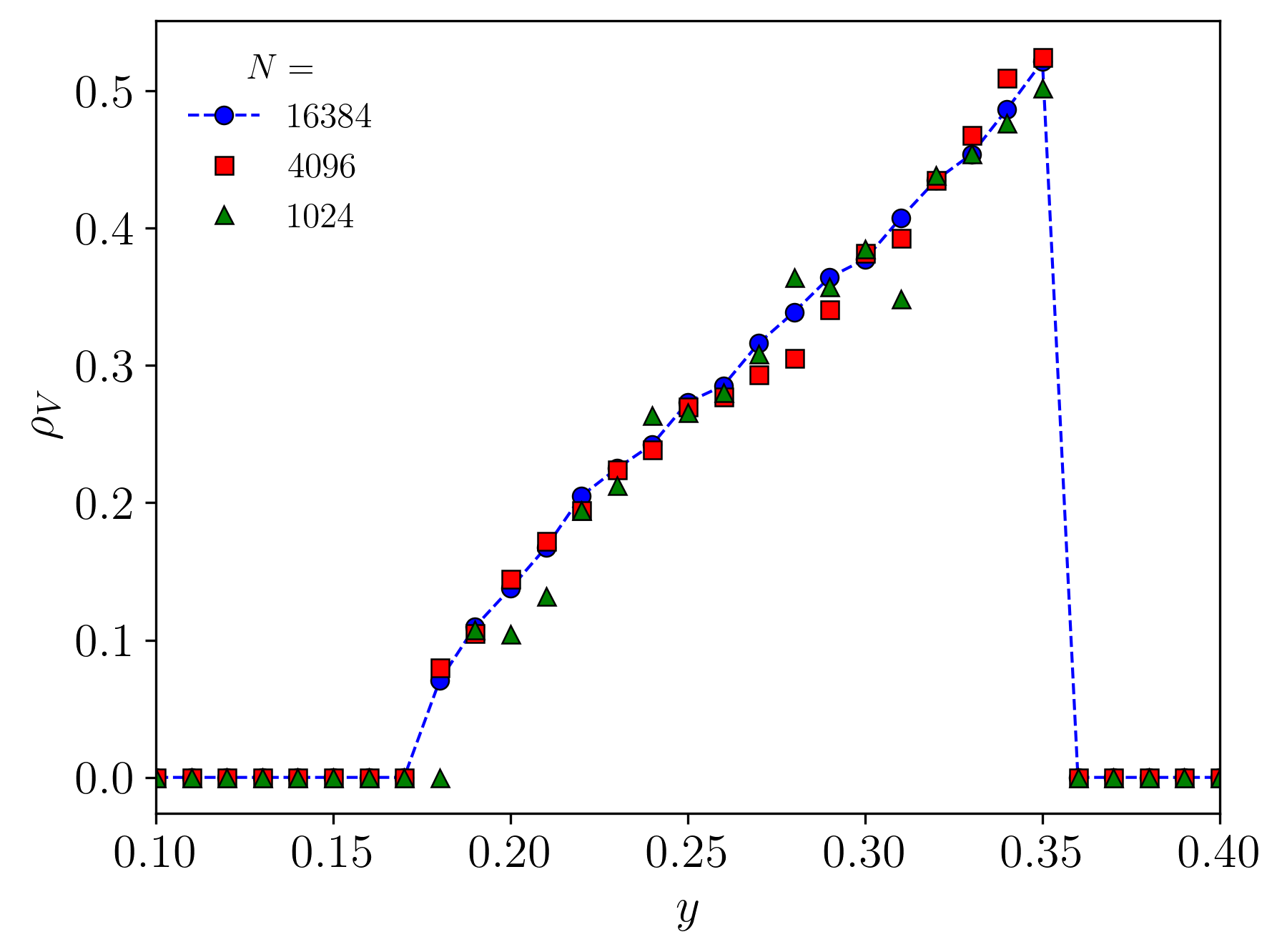}
\caption{Density $\rho_V$ as a function of $y$ for $N= 32^2,\ =64^2$ and $128^2$ with $\tau=5\times 10^6$ and $S=1$ for the HN.}
\label{fig_fss}
\end{center}
\end{figure}

Figure \ref{fig_fss} shows the density of vacant sites, $\rho_V$, as a function of the CO adsorption rate, $y$, for the hexagonal lattice, for $N=1024 =32^2,\ 4096= 64^2$ and $16384=128^2$.
In both cases, the densities were calculated after $\tau =5 \times 10^6$ MC steps and no average sampling ($S=1$). The idea is to identify the minimum size $N$ that yields reasonably stable results. This plot shows the active phase (with vacant sites and the consequent production of $\text{CO}_2$ and $\textbf{N}_2$ molecules), separating the two absorbing phases of the model (where $\rho_V=0$). As can be seen, the fluctuation of $\rho_V$ decreases as the number of sites increases. In addition, for $N=1024$, the critical point (the point that separates the first absorbing phase from the active one) is beyond those of the other two considered lattice sizes, which, in turn, are in agreement with the results found in the literature (for the HN). Although the lattices with $N=4096$ and $16384$ sites share the same second- and first-order transition points, we decided to use the larger one in order to minimize the fluctuation effects in our results. 

\begin{figure} [!htbp]
 \begin{center}
 \includegraphics[width = 2.2 in]{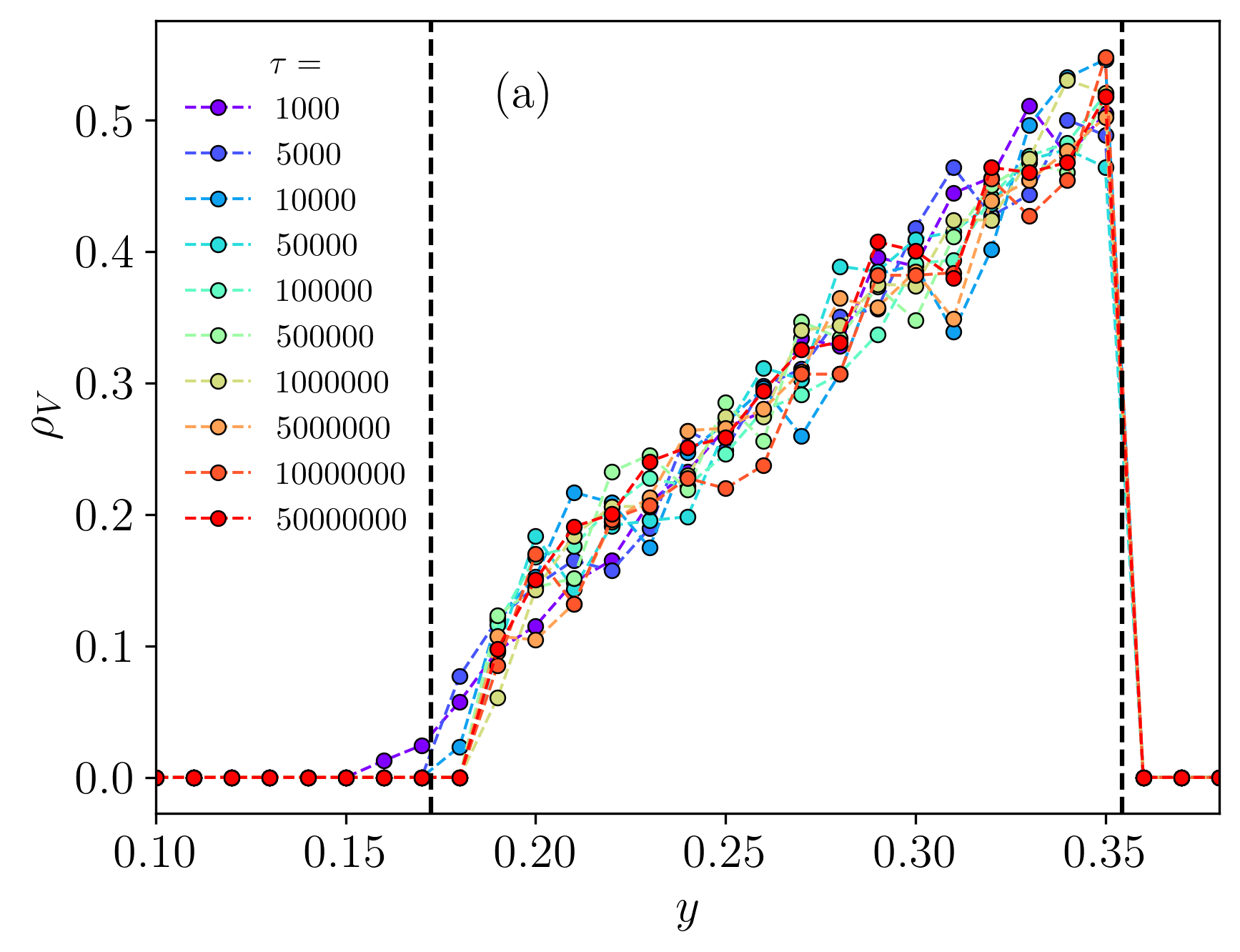}
 \includegraphics[width = 2.2 in]{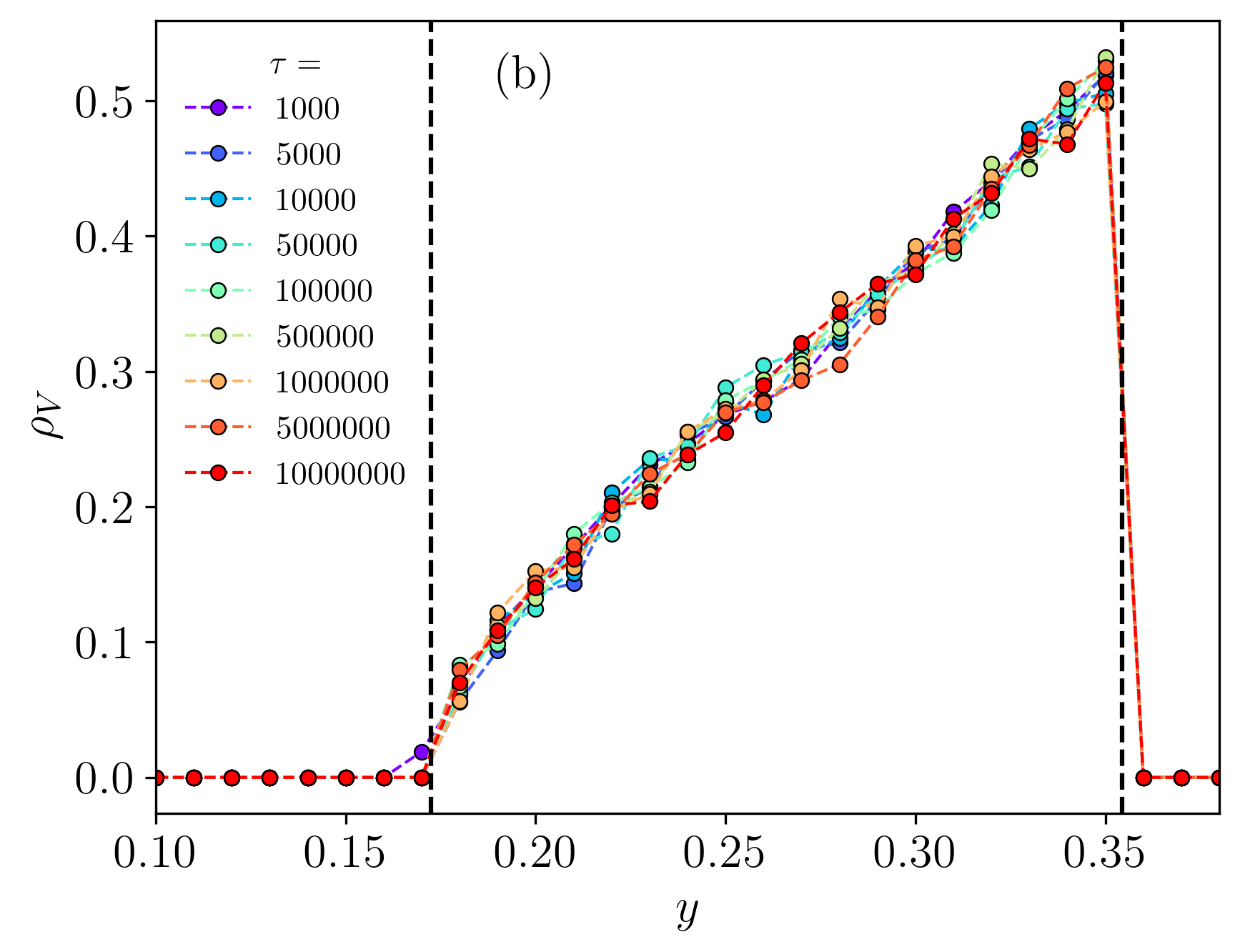}  \includegraphics[width = 2.2 in]{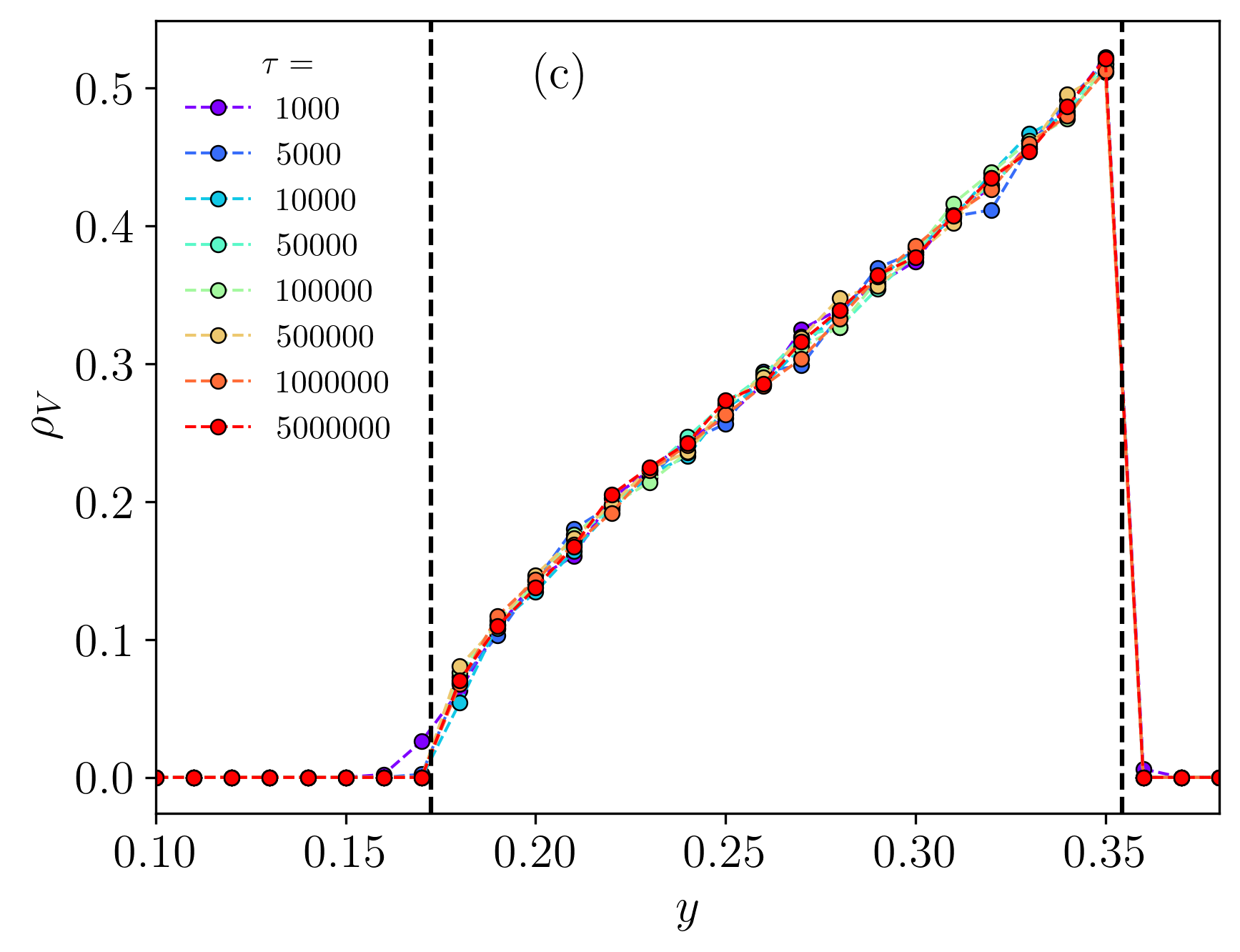}
 \caption{Analysis of $\tau$ for (a) $N=32^2$, (b) $N=64^2$, and (c) $N=128^2$ sites for the HN. The black dashed vertical lines are, respectively, the continuous- ($y_c = 0.1725$) and discontinuous-phase ($y_c = 0.3545$) transition points of the original model.}
 \label{fig_calibration_1}
 \end{center}
 \end{figure}

These fluctuations can also be seen in Fig. \ref{fig_calibration_1} which presents our analysis for the number $\tau$ of MC steps to be discarded before taking the averages. 
This figure shows two black dashed vertical lines separating the active phase, in the middle, from the two absorbing phases of the original model. The left vertical line delimits the continuous phase transition point, located at $y=y_1=0.1725(25)$, and the right vertical line, located at $y=y_2=0.3545(5)$ is related to the discontinuous phase transition point \cite{Brosilow1992}.

We performed simulations with $\tau$ ranging from $10^3$ to $5\times 10^6$, and, for $N=128^2$, we obtained reliable results for $\tau \geq 5\times 10^4$ MC steps. So, in order to ensure that the system reaches the steady-state regime, we decided to consider $\tau \geq 5\times 10^4$ MC steps in all of our simulations. 

We simulated the YK model on a hexagonal lattice for pedagogical purposes and for two main reasons: first, to calibrate $N$ and $\tau$; second, to obtain its phase diagram for comparison with the estimates we will derive for the two random networks considered in this work. 

\subsection{Final average degree}

The hexagonal lattice has $k_i = m = 6.0$ as average degree. However, the average degree $m$ follows a random distribution as function of $\mu$, for the ERN, or function of $r$, for the RGG. One different random instance of the network is generated for each $y$ value, so we have an average degree $m_i$ for each $y_i$. The final average network degree $K$ to be displayed on the results is the average of all $m_i$ values. Considering the ERN, we have $m \approx \mu$ and $K \approx \mu$. So we will display only $\mu$ for this network. However, for the RGG we have $K$ as a function of $r$. At the range of values of this work, the approximation $K(r) = \pi r^2$ has shown to be reasonable. Anyway, for the RGG we will show the $K$ value obtained directly from the average of all $m_i$ values. 

\subsection{Erdös-Rényi network}

\begin{figure} [!htbp] 
\begin{center}
\includegraphics[width = 3.0 in]{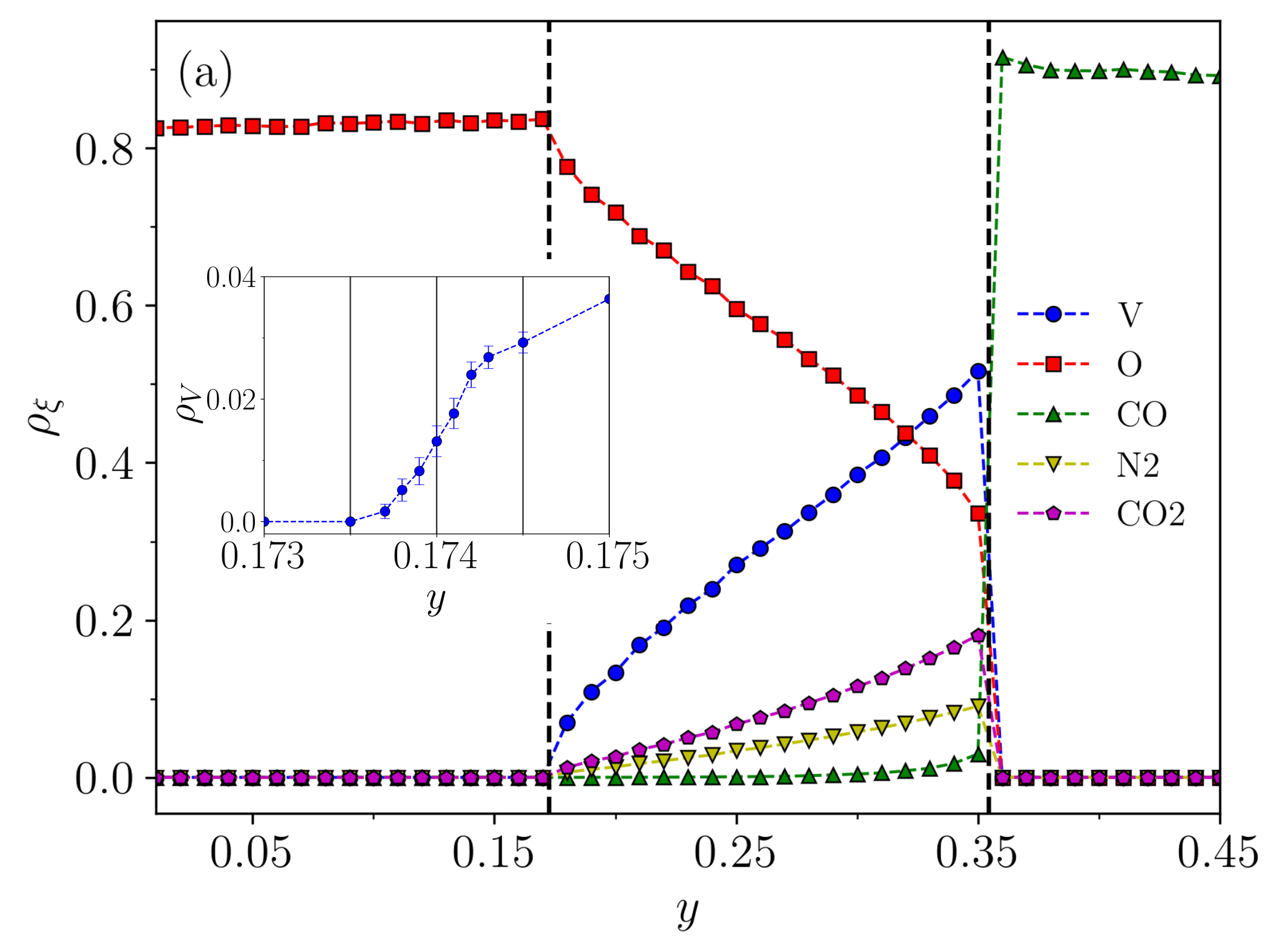}
\includegraphics[width = 3.0 in]{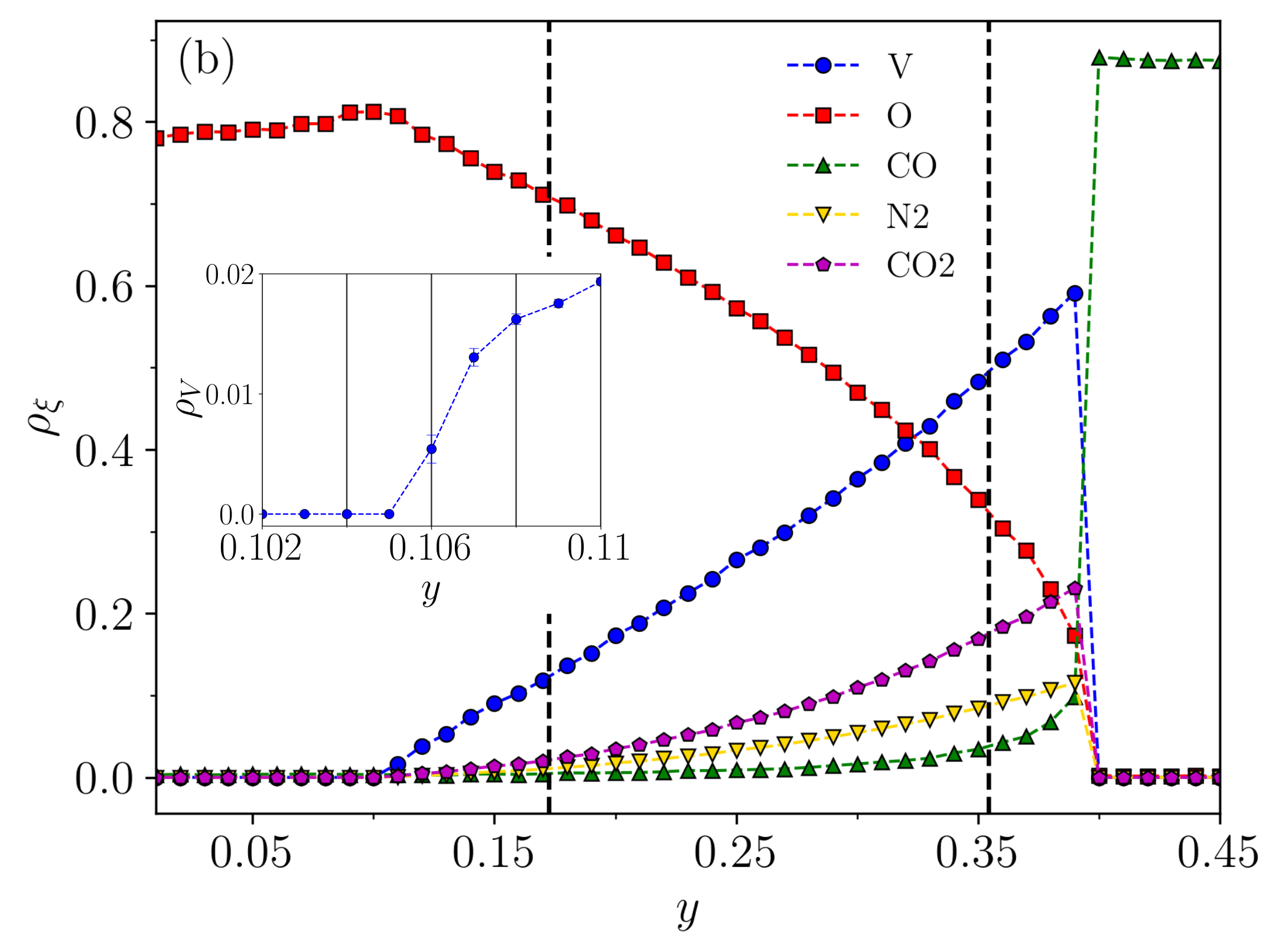}
\includegraphics[width = 3.0 in]{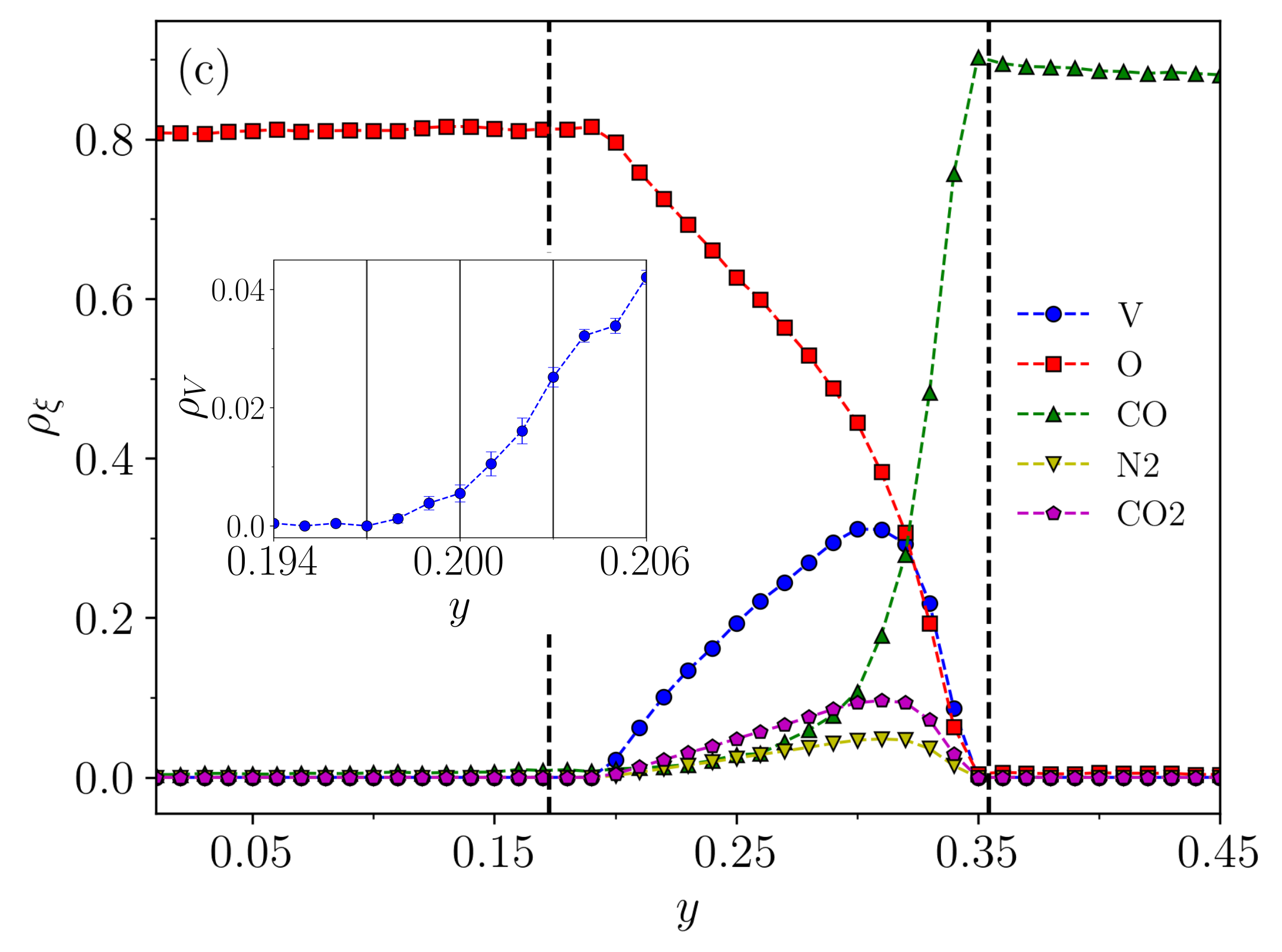}
\includegraphics[width = 3.0 in]{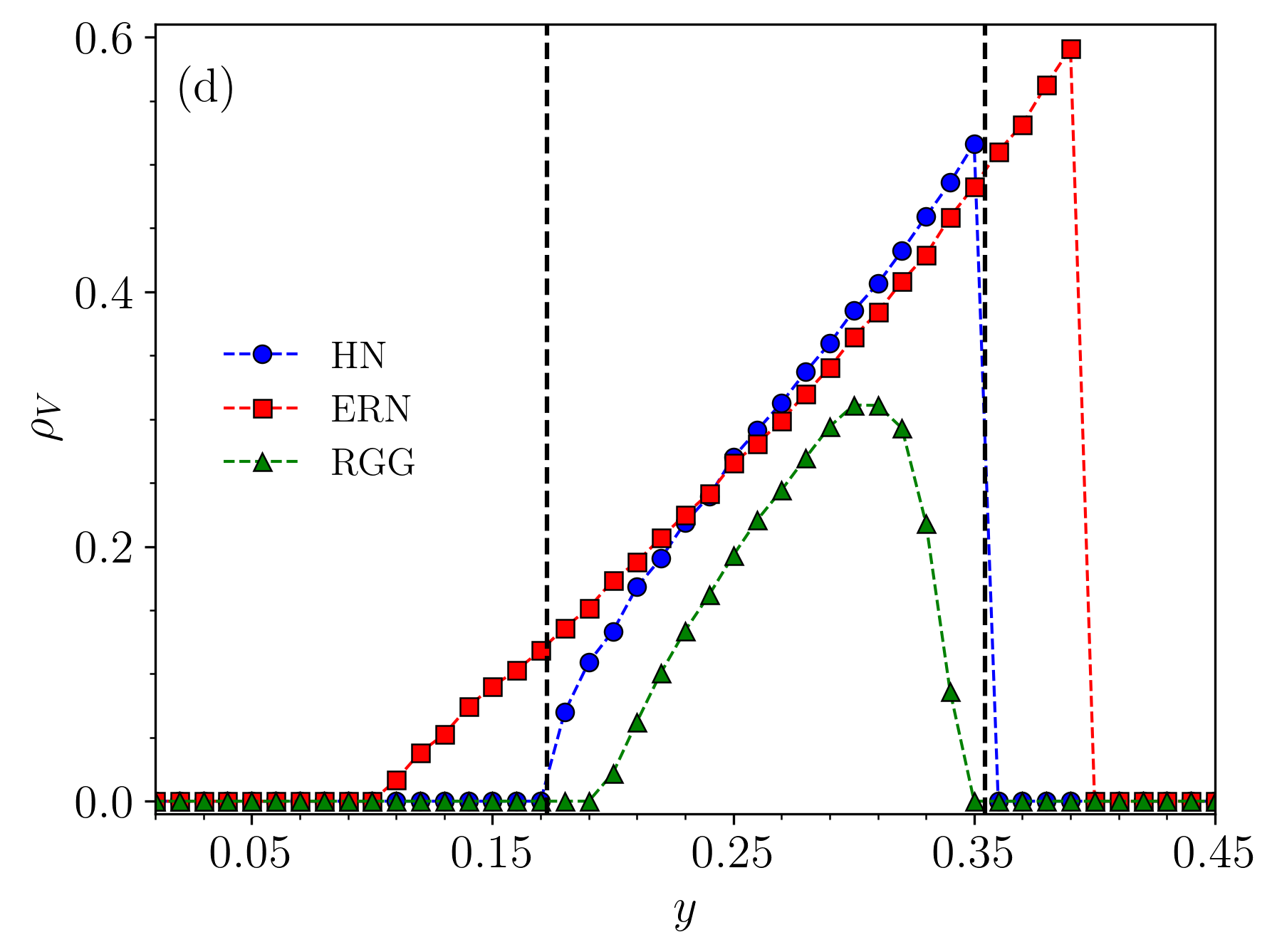}
\caption{Densities $\rho_{\xi}$ as function of $y$ for (a) Hexagonal network (HN), (b) Erdös-Rényi network (ERN), and (c) Random Geometric Graph (RGG). The insets show a zoom in on the second order transition and the error bars were obtained from 30 different time evolutions with different seed of the random number generator function. (d) Density of vacant sites, $\rho_V$, as function of $y$ for the three networks. The black dashed vertical lines indicate $y_1 = 0.1725$ and $y_2 = 0.3545$, the continuous and discontinuous phase transition points of the standard YK model. }
\label{fig_YK_s62g25}
\end{center}
\end{figure}

To enable the observation of both similarities and discrepancies in the phase diagrams of the three different lattices, we decided to simulate, at first, the two random networks with average degrees as close to six as possible, which means $\mu=6.0$ for the ERN. Considering the RGG, after an interpolation of $m$ for different values of $r$, we chose $r=1.38871$, yielding $K\simeq 6.06$. Figure \ref{fig_YK_s62g25} shows our results of the densities $\rho_{\xi}$ as function of $y$ for the (a) HN, as well as (b) ERN and (c) RGG networks. In addition, both figures present two black dashed vertical lines separating the absorbing phases to the active state of the standard YK model (Fig. \ref{fig_YK_s62g25}(a)). The inset of these figures show the region around the critical point $y_1$ (second order transition) of each model. 

Figure \ref{fig_YK_s62g25}(b) represents the phase diagram of the ERN for $\mu=6.0$ and shows that the continuous and discontinuous phase transitions remain, as in the original lattice, however, with a larger active phase window ($y_1\simeq 0.10$ and $y_2\simeq 0.40$). On the other hand, in Fig. \ref{fig_YK_s62g25}(c), which represents the phase diagram of the RGG, we observe that, although there also exists an steady reactive state separating two absorbing phases at $y_1\simeq 0.19$ and $y_2\simeq 0.35$, it is not clear that $y_2$ is a discontinuous phase transition point since the behavior of the densities as they approaches $y_2$ are smoother, as with a second-order transition. Although it is beyond the scope of the present work, this result suggests that, instead of presenting one continuous and one discontinuous phase transition for $r_{\text{NO}}=1.0$, as it happens with the regular hexagonal lattice and ERN, the RGG possesses two second-order phase transitions. More interestingly, if we consider that these transitions continue to exist for $r_{\text{NO}<1.0}$, we can obtain two lines of second-order phase transitions in $r_{\text{NO}\times y}$ space, for a given $K$, which eventually ends in a point where the reactive state disappears, placing this version of the YK model in a very restricted class of models that possess two lines of second order phase transitions, such as the ones presented in Refs. \cite{Bagnoli2001, Bagnoli2005}. Another interesting work, conducted by Ahmad and Balock \cite{Ahmad2007}, took into consideration the Eley-Rideal mechanism for the NO-CO reaction on a catalytic surface with inactive impurities, and showed that the discontinuous phase-order transition is converted into the continuous one.
In Fig. \ref{fig_YK_s62g25}(d), we have put together the density of vacant sites for the three networks to make clearer the similarities and differences of their phase transitions.

\begin{figure} [!htbp] 
\begin{center}
\includegraphics[width = 3.2 in]{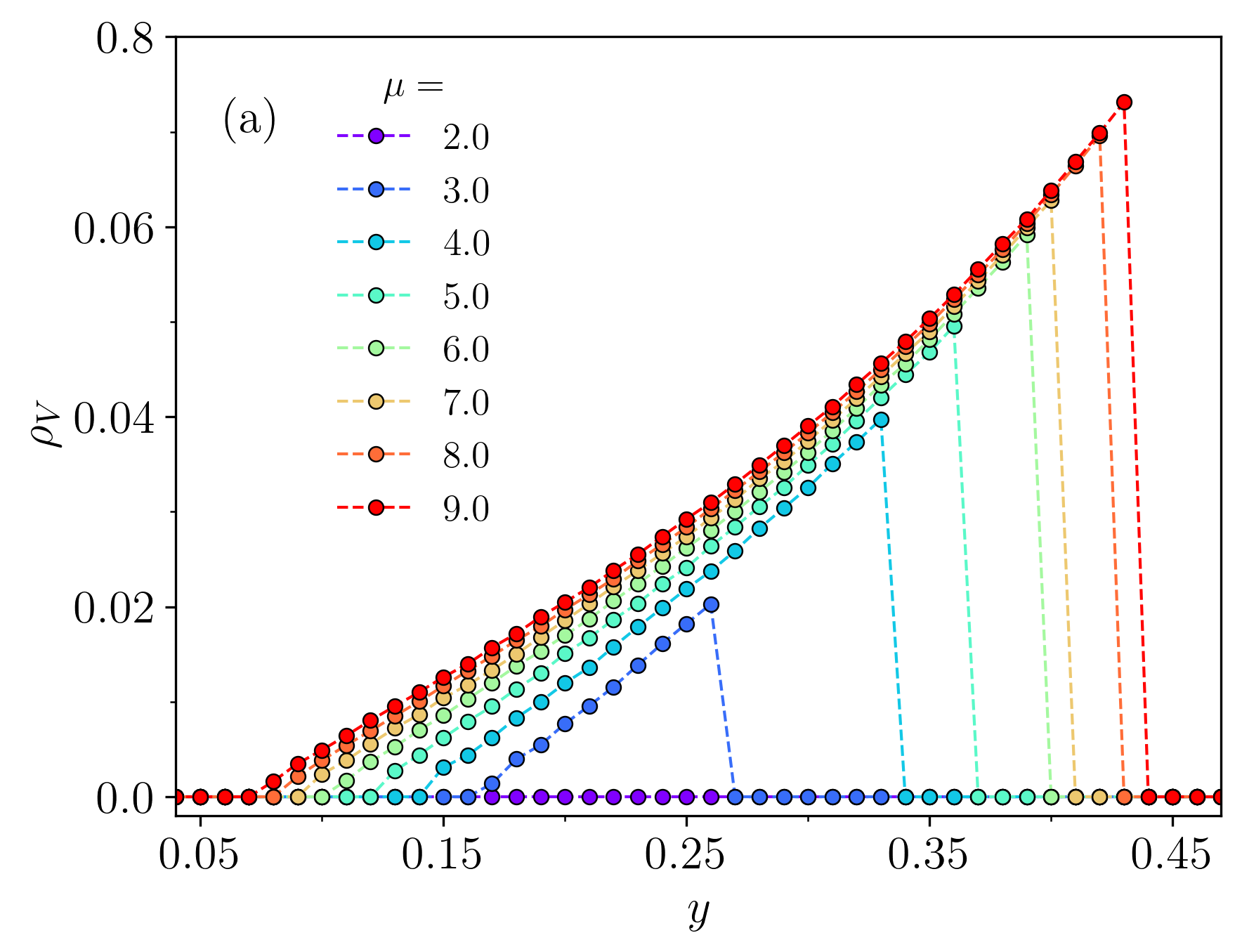}
\includegraphics[width = 3.2 in]{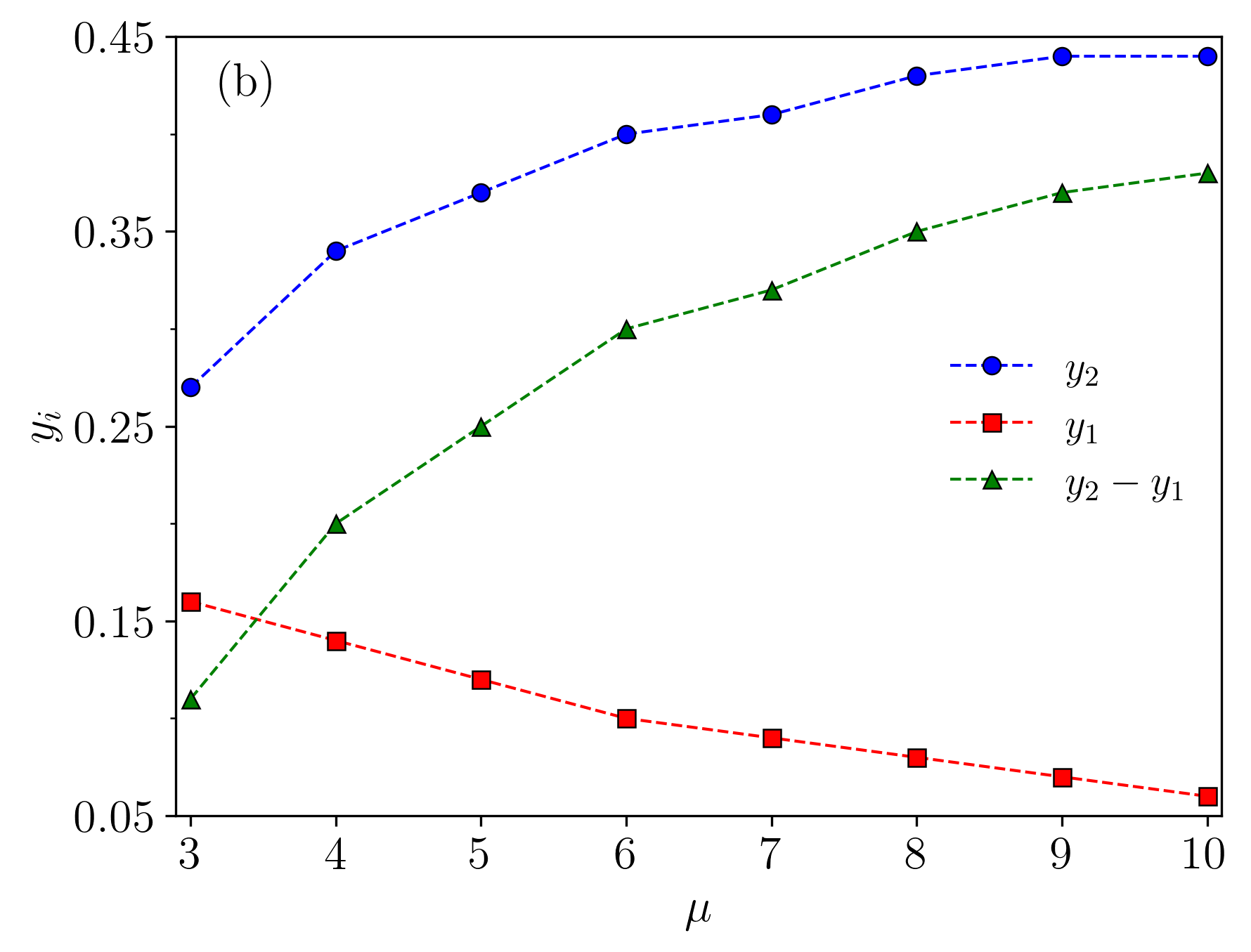}
\caption{Steady reactive window for the ERN with different values for the average degree $\mu$. (a) Density of vacant sites $\rho_V$ as function of $y$ for $2.0 \leq \mu \leq 9.0$. (b) Continuous ($y_1$) and discontinuous ($y_2$) phase transition points for the different values of $\mu$, and the reactive window ($y_2-y_1$) of the model. The error bars are smaller than the symbols.}
\label{fig_YK_s66g15}
\end{center}
\end{figure}

In the following, we look into the model for different values of the average degree in order to observe the behavior of the phase transitions and the steady-state reactive window for both random networks. Figure \ref{fig_YK_s66g15} presents our results for the ERN for $2.0 \leq \mu \leq 9.0$. 

As can be seen in Fig. \ref{fig_YK_s66g15}(a), the reactive window begins within the range $0.16 < y < 0.26$, and its width grows with increasing values of $\mu$.  This figure also shows that the character of the transitions remains unchanged throughout the range of values of $\mu$, showing that the order of the transitions is ``strong'' no matter how many neighbors each site has. In addition, the greater the average number of neighbors, the less likely the system is to reach one of the absorbing states. This behavior is shown in Fig. \ref{fig_YK_s66g15}(b): the red and blue dots represent, respectively, the continuous ($y_1$) and discontinuous ($y_2$) phase transition points as funcion of $\mu$, while the green dots represent the steady-state reactive window ($y_2-y_1$). The growth rate slows down, demonstrating that the width of the reactive window will eventually reach a constant value for larger values of $\mu$.

\begin{figure} [!htbp] 
\begin{center}
\includegraphics[width = 3.5 in]{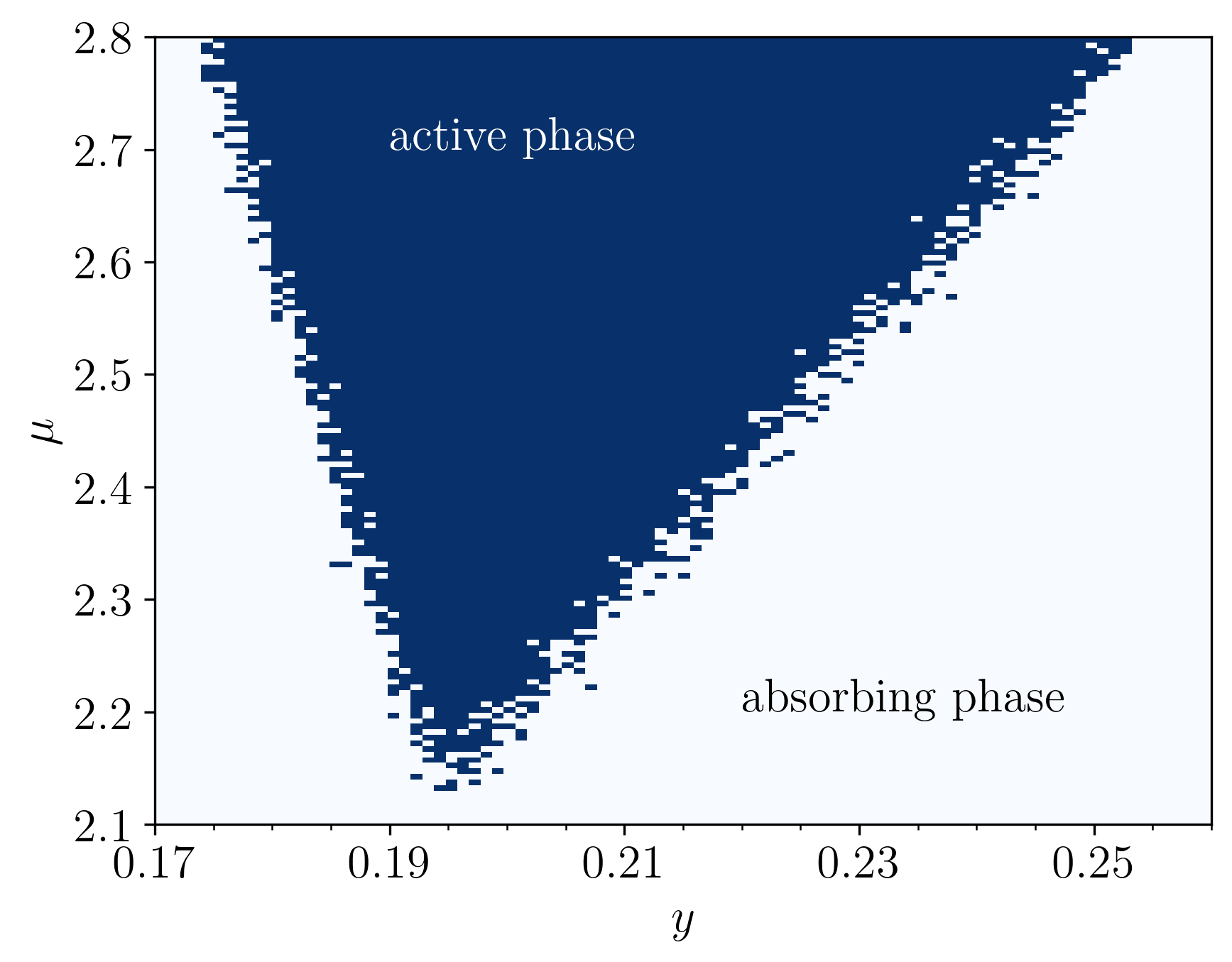}
\caption{Phase diagram in the $(y,\mu)$ space for the ERN. The blue color dots indicate the active phase, defined as the density of vacant sites greater than zero ($\rho_V > 0.0$), and the gray color dots indicate that the system is in the absorbing state ($\rho_V=0.0$). The points were obtained with $\Delta y = 0.001$ and $\Delta \mu = 0.005$.}
\label{fig_YK_s68g19}
\end{center}
\end{figure}

Although it is not the goal of the present work, we decide to look into the phase diagram of the model simulated in the ERN for small values of $\mu$, aiming to observe the beginning of the reactive phase. Figure \ref{fig_YK_s68g19} presents the result of this analysis in the $(y,\mu)$ space, obtained for $N=128^2$, $\tau=10^6$ and $S=10^3$.
The absorbing phase is obtained when the density of vacant sites is equal to zero ($\rho_V=0.0$). On the other hand, the system possesses reactive states for a given $\mu$ when $\rho_V > 0.0$ for some value of $y$. In this case, we observed that the steady-state reactive window begins for $\mu\approx 2.1$ and grows continuously from there on.

\begin{figure} [!htbp] 
\begin{center}
\includegraphics[width = 6.0 in]{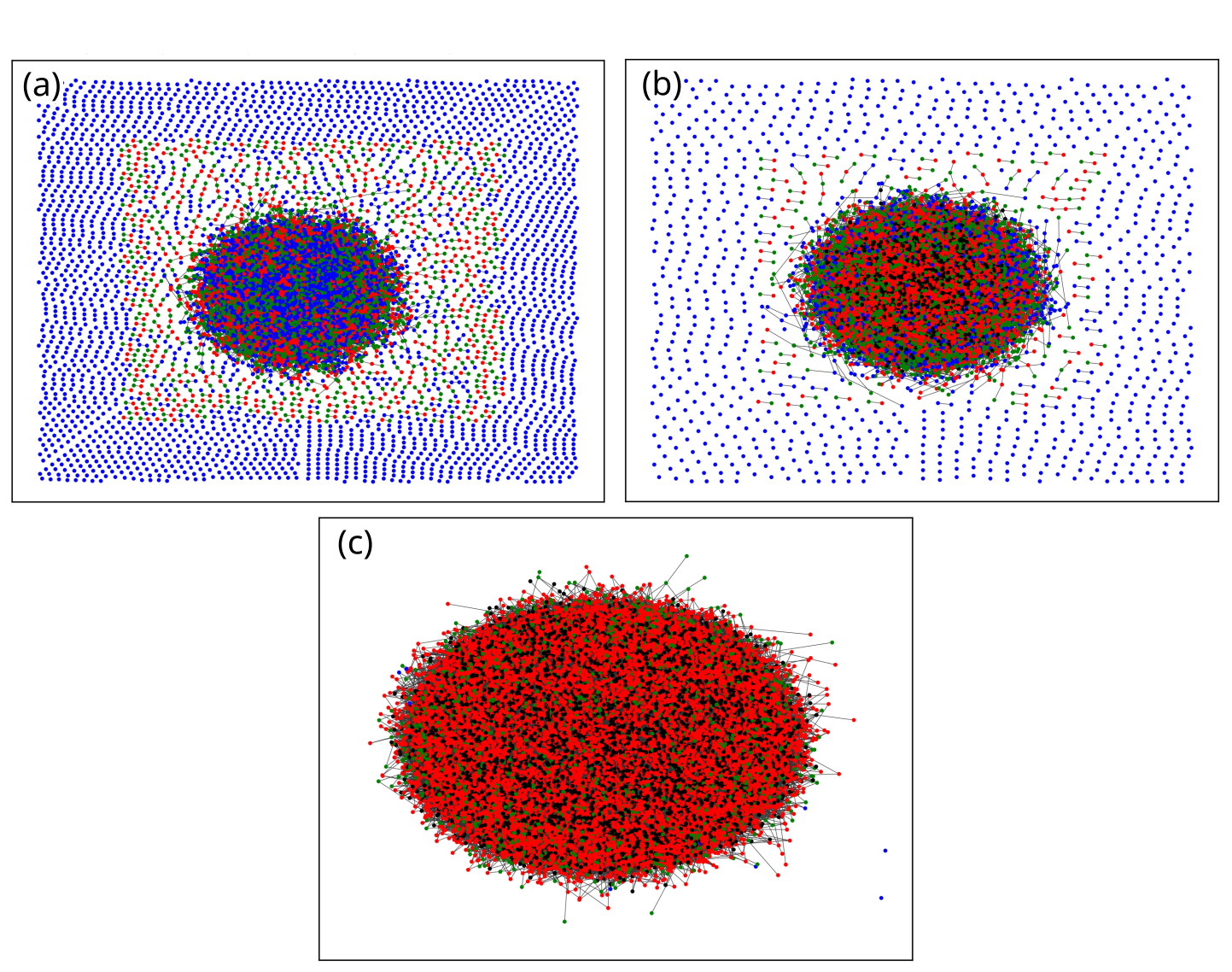}
\caption{Snapshot of the Erdös-Rényi network with different values of $\mu$. The largest component is located at the center of the plot and isolated nodes are located further away from it. Parameters: $N=128^2$, $\tau=10^6$, $S=10^3$ and $y=0.24$. Initial state corresponds to $t=0$ and all sites are vacant. Final state means a time evolution over $\tau + S$ MCS. Vacant sites are indicated in black (\CIRCLE), adsorbed CO molecules in blue ({\color{blue}\CIRCLE}), O atoms in red ({\color{red}\CIRCLE}), and N atoms in green ({\color{green}\CIRCLE}). The snapshots were taken after $\tau + S$ MC steps with $\tau=10^6$, $S=10^3$ and $y=0.24$. (a) $\mu = 2.0$: the system does not present steady-state active phase and $s_c \approx 0.79278$, $\rho_V = 0.0$, $\rho_{\text{CO}} \approx 0.49$, $\rho_{\text{O}} \approx 0.22$, and $\rho_{\text{N}} \approx 0.3$. (b) $\mu = 3.0$: the system already presents active phase and $s_c \approx 0.93386$, $\rho_V \approx 0.16$, $\rho_{\text{CO}} \approx 0.16$, $\rho_{\text{O}} \approx 0.4$, and $\rho_{\text{N}} \approx 0.27$. (c) $\mu = 9.0$: the system still presents active phase and $s_c \approx 0.99990$, $\rho_V \approx 0.27$, $\rho_{\text{CO}} = 0.0$, $\rho_{\text{O}} \approx 0.62$, and $\rho_{\text{N}} \approx 0.1$.}
\label{snapshots_ERN_v2}
\end{center}
\end{figure}

In Fig. \ref{snapshots_ERN_v2} we present snapshots for the ERN with three different average degrees, $\mu$. The configurations were calculated at $y=0.24$, which is around the middle point of the reactive window when $\mu = 3.0$ (see Fig. \ref{fig_YK_s66g15}(a)). This CO adsorption rate is located inside the reactive window for all values of $\mu$ considered in this work, except the lowest value which was purposely chosen because it does not present an active phase. 

Figure \ref{snapshots_ERN_v2}(a) shows the network for $\mu=2.0$ with its largest component located at the center of the plot and isolated sites located further away from it. By the largest component ($S_c$), we mean the largest group of interconnected sites within the network, and its density is simply $s_c=S_c/N$. When $s_c$ approaches 1.0, we have a percolated network. As shown above, the system does not present an active phase for $\mu=2.0$ and, in this case, the density of the largest component is $s_c\approx 0.79278$. This indicates a network entering the percolated phase \cite{Barabasi2016}. Figures \ref{snapshots_ERN_v2}(b) and (c) show, respectively, the snapshots of the network for $\mu=3.0$ and 9.0. In both cases, the system presents a steady-state active phase. For $\mu=3.0$ and $\mu=9.0$ one obtains $s_c\approx 0.93386$ and $s_c\approx 0.99990$, that is, both networks are percolated.

\subsection{Random Geometric Graph}

\begin{figure} [!htbp] 
\begin{center}
\includegraphics[width = 3.2 in]{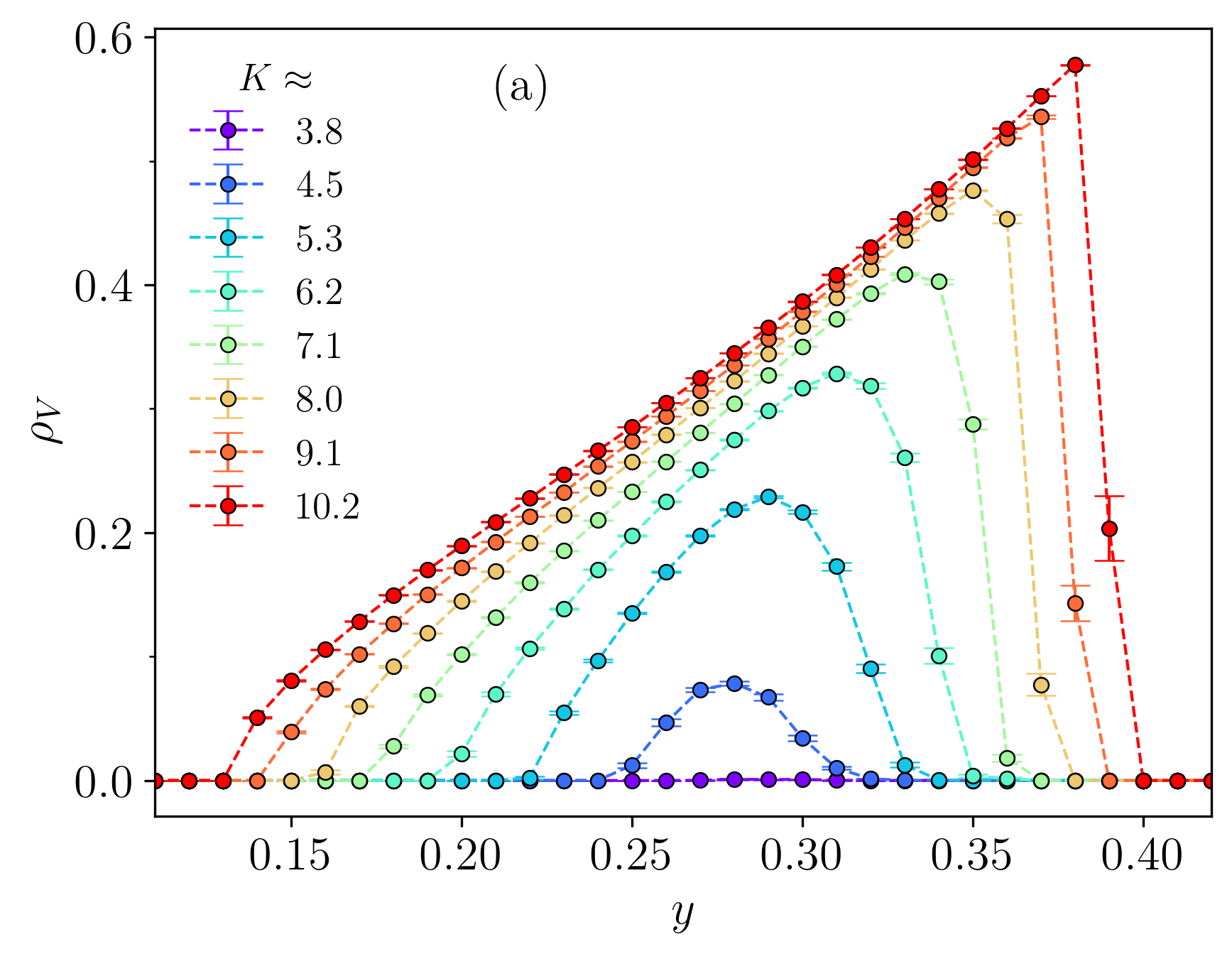}
\includegraphics[width = 3.2 in]{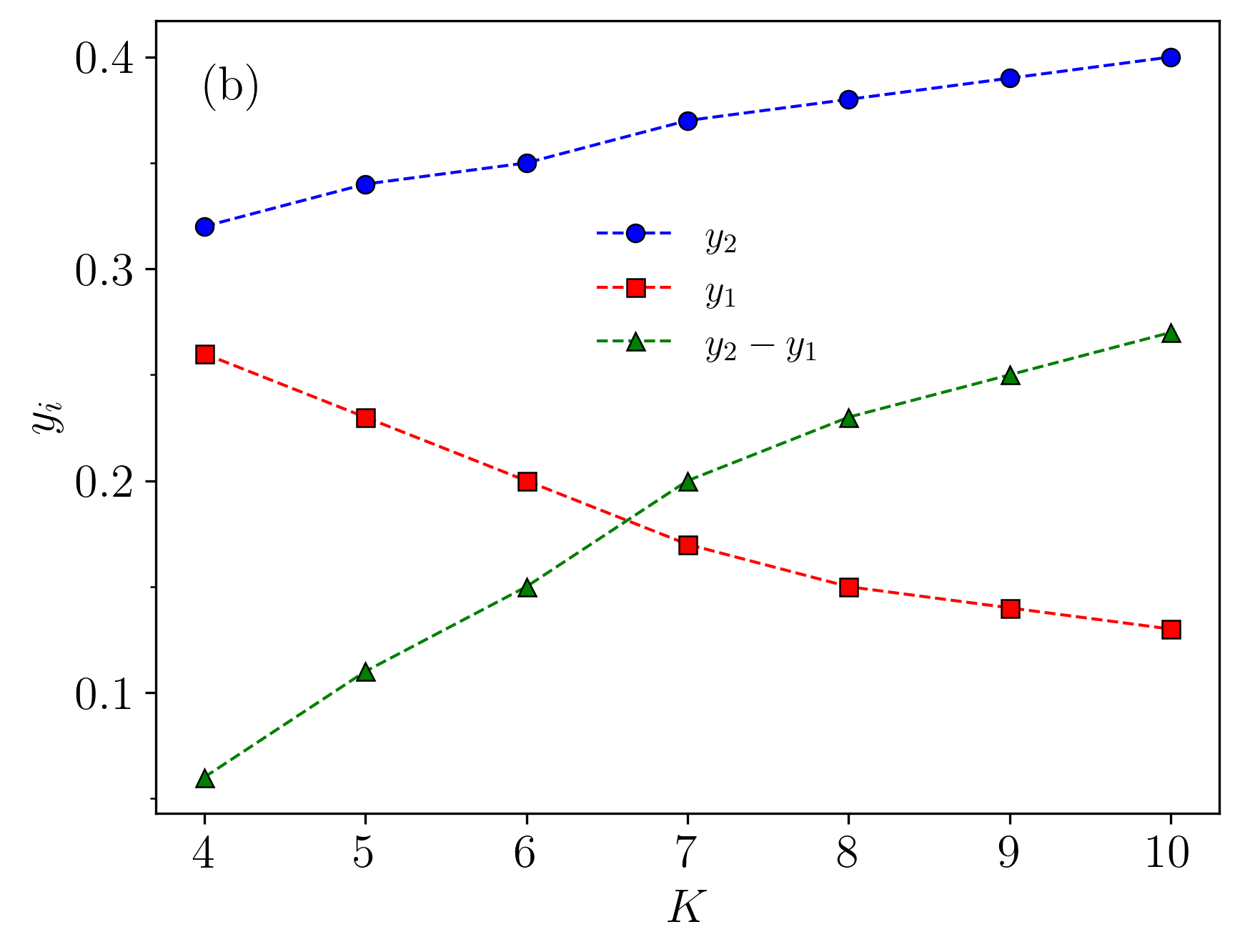}
\caption{Steady-state reactive window for the RGG with different values for the average degree $K$. Parameters: $N=128^2$, $\tau=10^6$, $S=10^3$ and $g=21$. (a) Density of vacant sites $\rho_v$ as function of $y$ for $4.0 \leq K \leq 10.0$. The error bars are the standard error of the mean $\delta = \sigma / \sqrt{g}$, where $\sigma$ is the standard deviation, and $g=21$ is the number of independent runs. (b) Phase transition points, $y_1$ and $y_2$, for the different values of $K$, and the reactive window ($y_2-y_1$) of the model.}
\label{fig_YK_s72g9}
\end{center}
\end{figure}

The same analysis was performed for the RGG, although it appears to be more challenging for our purpose in the present work. Figure \ref{fig_YK_s72g9}(a) shows the behavior of $\rho_V$ as a function of $y$ for $4.0 \leq K \leq 10.0$, and, different from the ERN, the reactive window starts between $3.8 < K < 4.5$ with both $y_1$ and $y_2$ being critical points. However, the phase transitions related to $y_2$ seem to change from continuous to discontinuous for increasing values of $K$, as observed mainly for $K \geq 9.1$, which brings with it the need for further investigation of the character of the transitions as well as their dependence on $r_{\text{NO}}$, which is beyond the scope of this paper. However, as stated above, a similar conversion of one phase transition order to another was observed when varying the impurities on a catalytic surface in NO-CO reactions \cite{Ahmad2007}. Figure \ref{fig_YK_s72g9}(b) presents the two transition points of the model ($y_1$ and $y_2$) as functions of $K$ along with the reactive window $y_2-y_1$ which also grows with increasing values of $K$.

\begin{figure} [!htbp] 
\begin{center}
\includegraphics[width = 2.2 in]{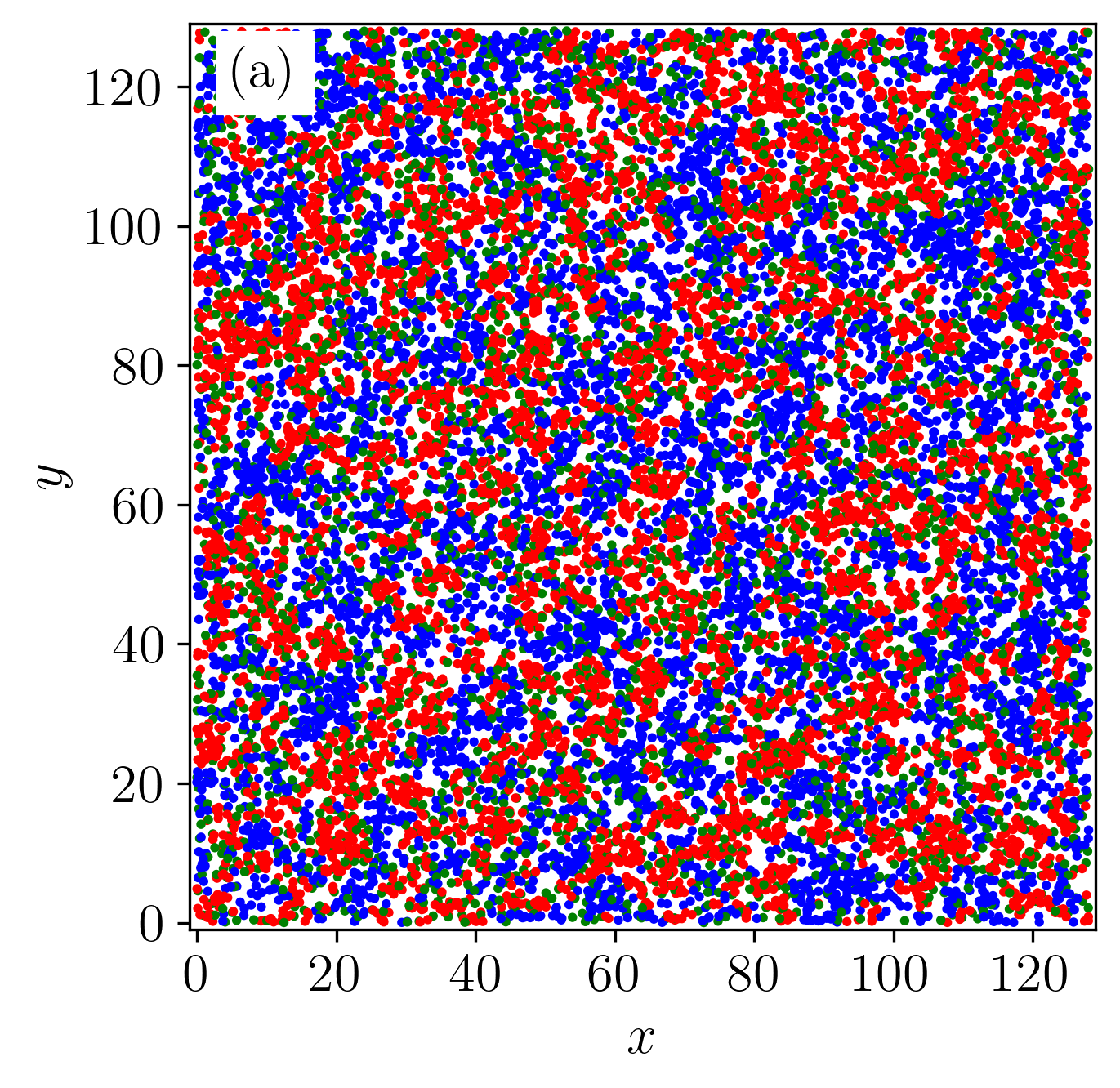}
\includegraphics[width = 2.2 in]{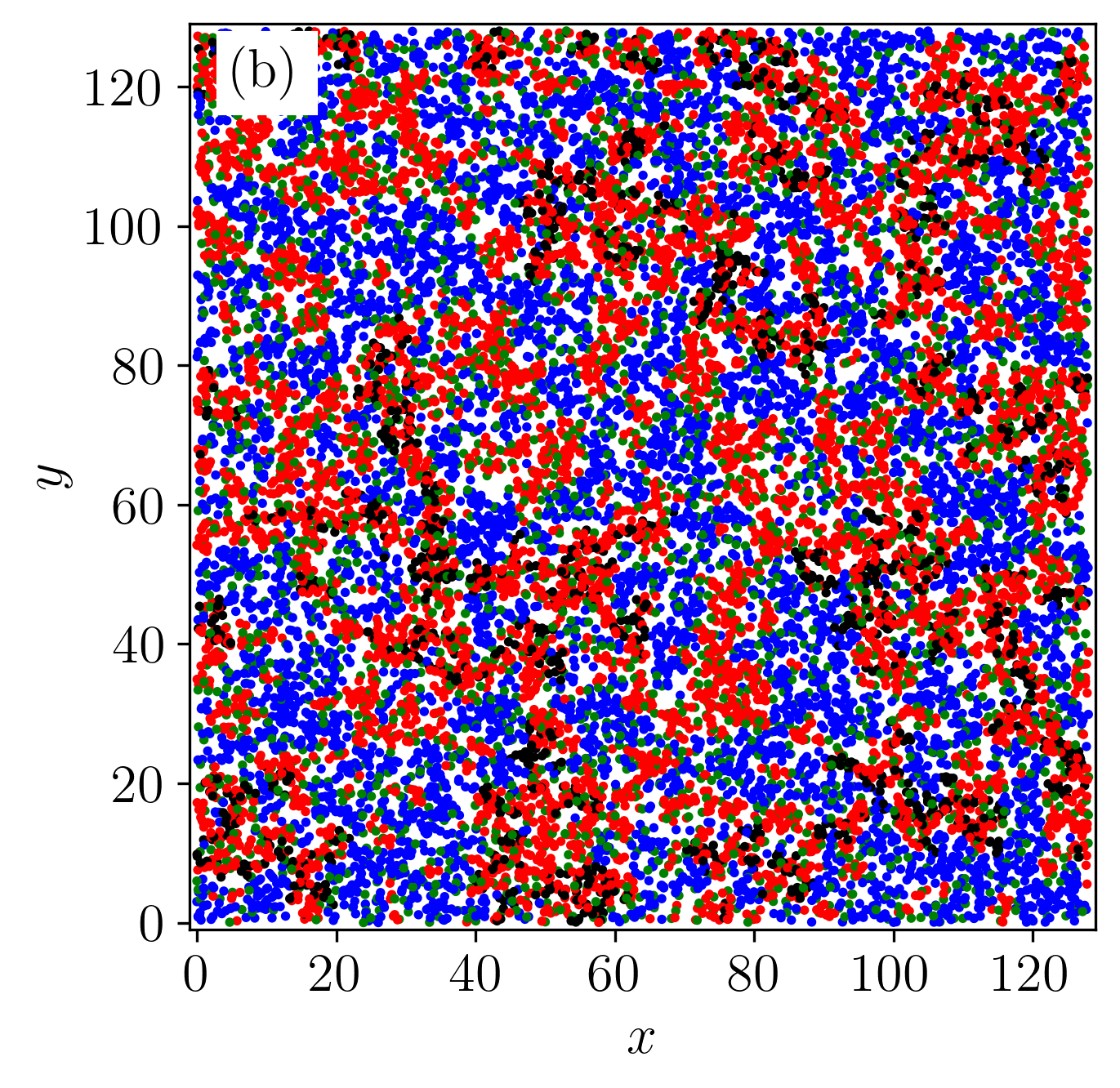}
\includegraphics[width = 2.2 in]{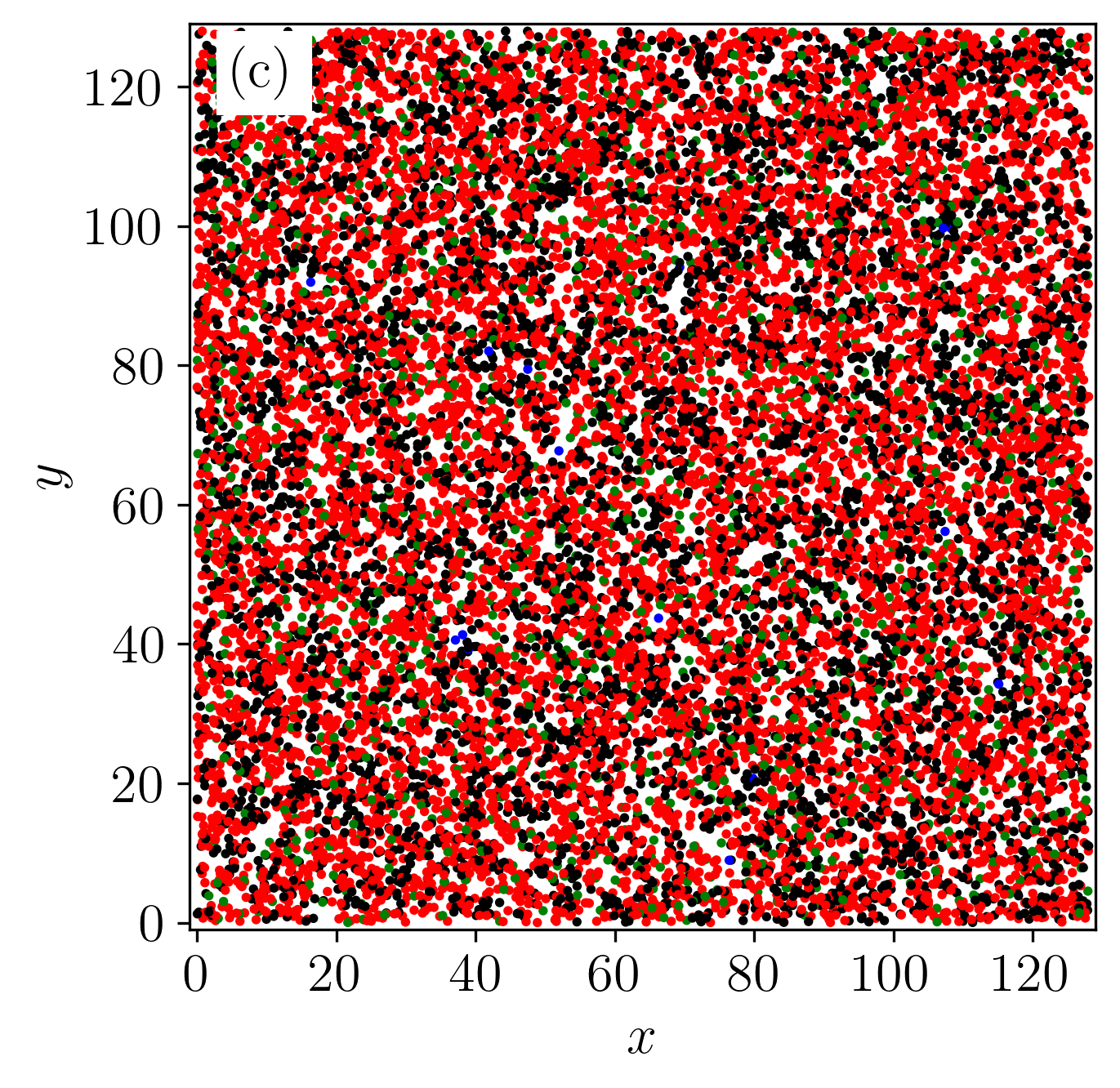}
\caption{Snapshot of the RGG with different average degrees $K$. Vacant sites are indicated in black (\CIRCLE), adsorbed CO molecules in blue ({\color{blue}\CIRCLE}), O atoms in red ({\color{red}\CIRCLE}) and N atoms in green ({\color{green}\CIRCLE}). Parameters: $N=128^2$, $\tau=10^6$, $S=10^3$ and $y=0.28$. (a) $r = 1.1$ and $K \approx 3.8$: the system does not present reactive window and $s_c \approx  0.05457$, $\rho_V = 0.0$, $\rho_{\text{CO}} \approx 0.36243$, $\rho_{\text{O}} \approx 0.40472$, and $\rho_{\text{N}} \approx 0.23285$. (b) $r = 1.2$ and $K \approx 4.5$: the system presents steady-state reactive state and $s_c \approx 0.81250$, $\rho_V \approx 0.09554$, $\rho_{\text{CO}} \approx  0.35030$, $\rho_{\text{O}} \approx 0.36748$, and $\rho_{\text{N}} \approx 0.18667$. (c) $r = 1.8$ and $K \approx 10.2$: The system possesses a large reactive window and $s_c \approx 0.99988$, $\rho_V \approx 0.34540$, $\rho_{\text{CO}} = 0.00116$, $\rho_{\text{O}} \approx  0.57648$, and $\rho_{\text{N}} \approx 0.07699$.}
\label{fig_YK_s75g15}
\end{center}
\end{figure}

The snapshots of the RGG taken for three different values of $K$ after $\tau=10^6$ MC steps are presented in Fig. \ref{fig_YK_s75g15}. The configurations were obtained for a catalytic surface with $N=128^2$ sites and for $y=0.28$. In Fig. \ref{fig_YK_s75g15}(a), $K\approx 3.8$ and, as can be seen, there are no vacant sites (black dots), meaning that the system does not present an active phase. The percolation transition for the RGG is around $r =1.0$. In this case, the density of the largest component is $s_c\approx 0.05457$, indicating that the network is not yet percolated, although it is in the critical region. Figure \ref{fig_YK_s75g15}(b) shows a snapshot for $k\approx 4.5$ which presents a small reactive window that allows the production of $\text{CO}_2$ and $\text{N}_2$, and consequently the presence of vacant sites. The largest component possesses a much higher value, $s_c\approx 0.81250$ while in Fig. \ref{fig_YK_s75g15}(c), $K\approx 10.2$ and $s_c\approx 0.99988$, indicating a percolated network on both cases.

\section{Conclusions \label{sec_conclusions}}

In this work we study the influence of random networks on the phase diagram and the nature of the phase transitions of the Yaldram-Khan (YK) model. We show that the Erdös-Rényi network preserves both continuous and discontinuous order transitions of the original model no matter the considered value of the average degree, $\mu$. The reactive window begins for a very small value of $\mu$ ( $\approx 2.1$), and increases with increasing $\mu$, as expected. On the other hand, the results for the random geometric graph show that the first-order phase transition is converted into second one for small values of the average degree, $K$. Only for large values of $K$ ($\approx 9.0$), the discontinuous transition is recovered. These findings indicate that randomness, whether long-range or local, significantly alters the phase diagram of the standard model. We think that these theoretical results may better reflect the inherent heterogeneity of real catalytic surfaces, and look forward to see future works focusing on experimental validation to confirm and refine these insights.

\section*{Acknowledgments}

This work was supported by the Brazilian agency CNPq (P. F. Gomes and H. A. Fernandes, grant no. 405508/2021-2; R. da Silva, grant no. 304575/2022-4) and by FAPEG (P. F. Gomes, grant no. 2019/10267000139). The research was conducted using the computational resources of LaMCAD/UFG. 

\appendix

\section{Computational details}

\begin{table} 
\centering
\begin{tabular}{c|c}
\hline \hline
Parameter &  Description    \\
\hline
\hline $K$ & measured network average degree \\
\hline $L$ & size of the square for the RGG network \\
\hline $\mu$ & input network average number of neighbors  \\
\hline $N$ & number of sites of the network   \\
\hline $r$ & RGG radius (control parameter) \\ 
\hline $\rho_V$ & density of vacant sites \\ 
\hline $S$ & number of MCS for the average calculation \\
\hline $t$ & time measured in MCS  \\
\hline $\tau$ & number of MCS in the thermalization process. \\
\hline $y$ & CO adsorption probability  \\
\hline
\hline
\end{tabular}
\caption{List of the used symbols on this work.}
\label{listofsymbols}
\end{table}

Table \ref{listofsymbols} shows the symbols used in this work for clarity purposes. The input parameters are $N$, $S$, $\tau$, $y$ and $\mu$ (for ERN) or $r$ (for RGG). The calculated average degree of the network is represented as $K$. For the ERN, we have $K \approx \mu$, while for the RGG $K$ is a numerical function of $r$. For the range of values we are using in this work, we have the approximated relation $K(r) \approx \pi r^2$ for the RGG. 

Each site can have 5 different states: empty (vacant), occupied by CO, occupied by O, occupied by N or occupied by NO. The initial condition is the empty network, where all sites are empty. The time evolution of the system comprises two stages and is performed over a total time of $t = \tau + S$ MC Steps. The first one is the thermalization one, which takes the system to the steady state. In this stage the densities start from their initial values and reach their final values over $\tau$ Monte Carlo steps (MCS), according to the system phase. The second stage is the sampling one, when the values of the densities are recorded to calculate their final average over time evaluation of $S$ MCSs. 

One MC step consists of $N$ analysis. The algorithm for one analysis is the following:
\begin{enumerate}
   \item The incident molecule is sorted: CO with chance $y$ and NO with probability $1-y$.
   \item A random site $s_1$ is selected. If it is occupied, nothing happens and a new analysis is initiated. The following steps happen if the site is vacant. 
   \item If a CO molecule is selected, it occupies the selected site $s_1$. Furthermore, if there is a neighboring oxygen-occupied site, there is a reaction creating CO2 (Eq. \ref{eq:co2_des}), which leaves the network. The two sites become vacant. The analysis ends and a new one is initiated.
   \item If a NO molecule is selected, it is dissociated in N and O (as we use $d=1.0$). If the selected site $s_1$ has neighbors, one of them is randomly selected and called $s_2$. If $s_2$ is empty, N and O enter the network. There are two possibilities with equal probabilities: (a) N occupies $s_1$ and O occupies $s_2$ or (b) vice versa.
   \item Considering the first possibility (a), once the nitrogen atom is adsorbed and before the oxygen one is adsorbed, nitrogen searches for nitrogen-occupied neighbors other than $s_2$. If it finds one, reaction \ref{eq:n2_g} occurs: N2(g) is produced and the two sites become empty. However, if it finds a NO-occupied site, reaction \ref{eq:n2_g_o_ad} occurs: N2(g) is produced and oxygen is left out in the previous NO-occupied site. Furthermore, this oxygen searches for an CO-occupied neighbor, if it finds, CO2(g) is created (reaction \ref{eq:co2_des}) and the sites become empty.
   \item Considering the oxygen absorbed on site $s_2$, it looks for an CO-occupied neighbor site. If this search is successful, reaction \ref{eq:co2_des} occurs producing CO2 and two involved sites become empty.
   \item The case of the second possibility (b) is equivalent.
\end{enumerate}

The overall algorithm is the following:
\begin{enumerate}
   \item Input parameters: $N$, $S$, $\tau$, $y$ and $\mu$ or $r$.
   \item An instance of the network is generated on a random basis. The average degree $K$ is calculated. 
   \item The states are reset to the initial states: all sites are empty.
   \item Steady state stage: time evaluation over $\tau$ MCS. The states are updated.
   \item Sampling stage over $S$ MCS. The averages of $\rho_{\xi}$ are calculated. 
\end{enumerate}
Considering the RGG, the used value of $K$ in the results is the average considering all the networks of all $y$ values.

.

\end{document}